\documentclass[aps,prd,preprint,showpacs,nofootinbib,amsmath,amssymb]{revtex4-1}
\usepackage{epsf}
\usepackage{color}
\usepackage{dcolumn}
\usepackage{bm}
\everymath{\displaystyle}
\begin{document}

\title{Relativistic quantum mechanics of a Proca particle in Riemannian spacetimes}

\author{\firstname{Alexander J.}~\surname{Silenko}}
\email{alsilenko@mail.ru} \affiliation{Research Institute for
Nuclear Problems, Belarusian State University, Minsk 220030, Belarus,\\
Bogoliubov Laboratory of Theoretical Physics, Joint Institute for Nuclear Research,
Dubna 141980, Russia}

\begin{abstract}
Relativistic quantum mechanics of a Proca (spin-1) particle in Riemannian spacetimes is constructed. Covariant equations defining electromagnetic
interactions of a Proca particle with the anomalous magnetic moment and the electric dipole moment in Riemannian
spacetimes are formulated. The relativistic Foldy-Wouthuysen transformation with allowance for terms proportional
to the zero power of the Planck constant is performed. The Hamiltonian obtained agrees with the corresponding Foldy-Wouthuysen Hamiltonians derived for scalar and Dirac particles and with their classical counterpart. The unification of relativistic quantum mechanics in the Foldy-Wouthuysen representation is discussed.
\end{abstract}

\pacs{04.62.+v, 03.65.Pm, 04.20.Cv, 11.10.Ef}
\maketitle

\section{Introduction}

We present a general quantum-mechanical description of a Proca (spin-1) particle in Riemannian spacetimes. Its electromagnetic interactions are analyzed. The anomalous magnetic moment (AMM) and the electric dipole moment (EDM) are taken into account. The Foldy-Wouthuysen (FW) transformation
\cite{FW} is performed for a nonrelativistic Dirac particle. We obtain exact expressions for terms proportional to the zero power of the Planck constant. For this purpose, we apply the relativistic FW transformation method
developed and substantiated in Refs. \cite{JMP,relativisticFW,PhysRevAexponential}.
The use of the relativistic FW transformation allows one to express the relativistic quantum mechanics (QM) in the Schr\"{o}dinger form.

Various properties and applications of the FW transformation have been considered in Refs. \cite{BD,Urban,dVFor}.
The FW transformation is widely used in electrodynamics
\cite{JMP,electrodynamics,RPJ}, quantum field theory
\cite{neznamovEChAYa}, optics \cite{Reuse,Khan,ultrafast},
condensed matter physics \cite{condensed}, nuclear physics
\cite{nuclear,nuclone}, gravity \cite{PRD,gravity,Gosgrav}, the
theory of the weak interaction \cite{weak} and also quantum
chemistry \cite{ReiherWolfBook,ReiherTCA,local,NakajimaH,ReiherRev}.
It is applicable not only for Dirac fermions but also for
particles with any spins
\cite{Case,Bryden,Pursey,Khan,Guertin,Leon,YB,TMP2008,SpinunitEDM,Honnefscalar,PRDexact}.
Recently, the FW transformation has been successfully employed \cite{hiddensupersymmetry} to clarify the origin of the hidden supersymmetry
and superconformal symmetry \cite{paperhid} in some purely bosonic quantum systems.

In precedent studies of Proca quantum mechanics, a detailed analysis of electromagnetic interactions of a spin-1 particle has been based on the approach developed in Ref. \cite{YB}. The relativistic FW Hamiltonian of a spin-1 particle with the AMM has been derived in Ref. \cite{SpinunitEDM}. However, all precedent
investigations using the FW transformation \cite{YB,SpinunitEDM,Energy1,PRDexact,Spinunit} have been fulfilled in the framework of special relativity. We can also mention an analysis of QM of a Proca particle in the
Minkowski space made in Ref. \cite{ZamaniMostafazadeh}. Some studies of QM of a Proca particle in curved spacetimes have been carried out in Refs. \cite{Seitz,Spinosa,Bagrov,Dereli,Obukhov,Schambach}. In these works, the Cartan spacetime torsion has also been considered. The nonmetricity in Einstein-Proca solutions has been studied in Ref. \cite{Dereli}.
In Refs. \cite{Seitz,Spinosa,Bagrov}, Lagrangians of a Proca particle in Riemann-Cartan spacetimes have been obtained. The corresponding Proca equations excluding electromagnetic interactions have been presented in Ref. \cite{Spinosa}. The Proca equations with an inclusion of electromagnetic interactions have been obtained in Ref. \cite{Bagrov}. However, the FW transformation has not been used in Refs. \cite{Seitz,Spinosa,Bagrov,Dereli,Obukhov,Schambach}. The Wentzel-Kramers-Brillouin approximation and the quasiclassical trajectory-coherent approximation have been applied in Refs. \cite{Seitz,Spinosa} and in Ref. \cite{Bagrov}, respectively. In the present work, we demonstrate the possibility to obtain the classical limit of Proca QM for a relativistic spin-1 particle in strong electromagnetic and gravitational fields. For this purpose, we perform the subsequent Sakata-Taketani \cite{SaTa} and FW transformations. In the FW representation, the passage to the classical limit
usually reduces to a replacement of the operators in quantum-mechanical Hamiltonians and equations of motion with the
corresponding classical quantities \cite{Classlimit}. Previously, a detailed quantum-mechanical description of a scalar particle in Riemannian
spacetimes has been fulfilled in Ref. \cite{Honnefscalar}. For a Dirac particle, the corresponding problem has been solved in Refs. \cite{PRD,gravity}.

Our notations correspond to Refs. \cite{Honnefscalar,PRDThomas}. We denote world and spatial indices by Greek and Latin letters
$\alpha,\mu,\nu,\ldots
=0,1,2,3,~i,j,k,\ldots=1,2,3$, respectively. Tetrad
indices are denoted by Latin letters from the beginning of the
alphabet, $a,b,c,\ldots = 0,1,2,3$. 
Temporal and spatial tetrad indices are
distinguished by hats. The signature is $(+---)$.
Commas and semicolons before indices denote partial and covariant derivatives, respectively.
Repeated Greek indices and the Latin indices from the beginning of the
alphabet are summed over the values $0, 1, 2, 3$. Repeated Latin indices $i,j,k,\ldots,\widehat{i},\widehat{j},\widehat{k},\ldots$ are summed over the values $1, 2, 3$. The tetrad indices are raised and lowered with the flat Minkowski metric,
$\eta_{ab} = {\rm diag}(1,-1, -1, -1)$.

We use the system of units $\hbar=1,~c=1$ but include $\hbar$ and $c$ explicitly when this inclusion clarifies the problem.

\section{Comparison of quantum mechanics of Dirac and Proca particles}

A comparison of basic quantum-mechanical equations for Dirac and Proca particles is instructive for a consideration of fundamentals of Proca QM.

\subsection{Fundamentals of Dirac quantum mechanics}\label{DiracMinkowski}

Dirac QM describes a single spin-1/2 particle in the Minkowski spacetime. The action and the Lagrangian of spinor field are given by
\begin{equation} S=\int{{\cal L}d^4x},\qquad {\cal L}=\bar{\Psi}(i\gamma^\mu \partial_\mu -m)\Psi, \label{LagrangDir} \end{equation}
where $\bar{\Psi}=\Psi^\dag\gamma^0$.

The Euler-Lagrange equation reads
\begin{equation} \partial_\mu\frac{\partial{\cal L}}{\partial(\partial_\mu\bar{\Psi})}=\frac{\partial{\cal L}}{\partial\bar{\Psi}}=0.\label{EulerLaGrDir} \end{equation}

Explicitly, we obtain the Dirac equation for a free particle:
\begin{equation}\left(i\gamma^\mu \partial_\mu- m\right)\Psi=0. \label{Dirac0}\end{equation}

The $4\times4$ Dirac matrices satisfy the so-called Clifford algebra,
\begin{equation}\left\{\gamma^\mu, \gamma^\nu\right\}=2\eta^{\mu\nu}. \label{Diracma}\end{equation}
The fifth matrix $\gamma^5$ can also be introduced:
\begin{equation}\gamma^5=i\gamma^0\gamma^1\gamma^2\gamma^3, \quad \left(\gamma^5\right)^2=1,\quad
\left\{\gamma^5, \gamma^\mu\right\}=0. \label{gam5}\end{equation}

The Lagrangian of a Dirac particle in an electromagnetic field should be invariant under the local gauge transformation $\Psi\rightarrow\Psi'=\exp{(ie\Lambda)}\Psi$. It is well known that this condition results in a replacement of the partial derivative with the covariant (lengthened) one,
\begin{equation}
D_\mu = \partial _\mu + ieA_\mu,
\label{Dmu}
\end{equation} where $A_\mu$ is the four-potential of the electromagnetic field. We use the denotations $A_\mu=(\Phi,-\bm A)$, $\partial _\mu=(\partial _0,\nabla)$, where $\nabla$ is the nabla operator. The gauge transformations of the electromagnetic field have the form \begin{equation}
A_\mu\rightarrow A'_\mu= A_\mu-\partial _\mu\Lambda,\qquad\bm A\rightarrow\bm A'=\bm A+\nabla\Lambda,\qquad \Phi\rightarrow\Phi'=\Phi-\partial_0\Lambda. \label{gaugetf} \end{equation}
The Lagrangian and the Dirac equation in the electromagnetic field are given by
\begin{equation} {\cal L}=\bar{\Psi}(i\gamma^\mu D_\mu -m)\Psi, \label{LagrDirem} \end{equation}
\begin{equation}\left(i\gamma^\mu D_\mu- m\right)\Psi=0. \label{Diracem}\end{equation}
The Lagrangian of the electromagnetic field reads
\begin{equation} {\cal L}^{(em)}=-\frac14F_{\mu\nu}F^{\mu\nu}, \qquad F_{\mu\nu}=\partial_{\mu}A_{\nu}-\partial_{\nu}A_{\mu}. \label{Lagremf} \end{equation}

The Dirac equation in curved spacetimes can also be obtained with the Lagrangian approach (see, e.g., Ref. \cite{Gasperini}).
The gauge invariance under local transformations can
be restored by replacing the partial derivative $\partial_\mu$ with the gauge covariant derivative $D_\mu$. One needs to apply a flat tangent space defined by the tetrad of four-vectors $e_a^\mu$ satisfying the relation
$\eta^{ab}e_a^\mu e_b^\nu=g^{\mu\nu}$. The covariant generalization of the integration, $d^4x\rightarrow \sqrt{-g}\,d^4x,\,g=\det{g_{\mu\nu}}$, should be used. The spinor field $\Psi$ locally defined in the flat
tangent space does not have curved Riemann indices, and its total covariant derivative reduces to a Lorentz covariant derivative (see Refs. \cite{Gasperini,HLM})
\begin{equation}\begin{array}{c}
\partial_\mu\rightarrow D_\mu=\partial_\mu+\frac i4\sigma^{ab}\Gamma_{\mu ab},\qquad \sigma^{ab}=\frac i2\left(\gamma^a\gamma^b-\gamma^b\gamma^a\right),\\
\Gamma_{\mu ab} =-\Gamma_{\mu ba} = e_\mu^c \Gamma_{cab},\qquad \Gamma_{cab}=e_b^\mu e_c^\nu e_{a\mu;\nu}=\frac12\left(-C_{cab}+C_{abc}-C_{bca}\right), \\
C_{abc}=-C_{bac}=e_a^\mu e_b^\nu(e_{c\nu,\mu}-e_{c\mu,\nu}).
\end{array} \label{Lorcovdv} \end{equation} The anholonomic components of the connection $\Gamma_{\mu ab}$ are often called the Lorentz connection coefficients. In Eq. (\ref{Lorcovdv}), $C_{abc}$ are the Ricci rotation coefficients and $\sigma^{ab}$ are the generators of the local Lorentz
transformations of the spinor field: $$\Psi\rightarrow\Psi'=\exp{\left(-\frac i4\omega_{ab}\sigma^{ab}\right)}\Psi.$$

As a result, the action is given by \cite{Gasperini}
\begin{equation} S=\int{\bar{\Psi}(i\gamma^a D_a -m)\Psi\sqrt{-g}\,d^4x},\qquad  D_a=e_a^\mu D_\mu,\label{LagrangDirG} \end{equation} and the covariant Dirac equation reads
\begin{equation}\left(i\gamma^a D_a- m\right)\Psi=0. \label{Dirac1}\end{equation}
When the electromagnetic fields are taken into account, the covariant derivative takes the form (see Ref. \cite{gravity})
\begin{equation}
D_\mu=\partial_\mu+ieA_\mu+\frac i4\sigma^{ab}\Gamma_{\mu ab}.
\label{Lorcopem} \end{equation}

The Dirac equation for the spin-1/2 particle can also be obtained by
quantization of the corresponding classical system. The spin degrees of freedom are described with the Grassmann variables \cite{Grassmannalgebra,Beresin}. In such a
description, the model possesses a local supersymmetry controlling the introduction of particle interactions \cite{Beresin}.
Namely, the odd constraint generates (via the Poisson bracket) the even constraint being
a classical analog of the Klein-Gordon equation. The simple supersymmetric structure of the algebra with
one even and one odd constraint allows one to modify the odd
constraint to introduce the interaction. The modified even constraint
is generated via the Poisson bracket of the modified odd constraint
with itself, and the modified odd and even constraints will
commute (relative to the Poisson bracket) as in the free case. The mechanism has been
presented in Ref. \cite{anyon}.

We can mention the existence of bosonic symmetries of the Dirac equation \cite{Simulik}.

\subsection{Fundamentals of Proca quantum mechanics}\label{ProcaMinkowski}

Proca QM describes a single spin-1 particle, and a massive Proca particle has three independent components of the spin. The Proca Lagrangian takes the form
\begin{equation} {\cal L}=-\frac12U_{\mu\nu}^\dag U^{\mu\nu}+m^2U_\mu^\dag U^\mu, \label{LagrangianBc} \end{equation}
where $U_{\mu\nu}=-U_{\nu\mu}$ is defined by
\begin{equation} U_{\mu\nu}=\partial_\mu U_\nu-\partial_\nu U_\mu. \label{ProcaII} \end{equation}
The complex functions $U^\mu$ are used. The additional conditions
\begin{equation} \partial_\mu U^\mu=0,\qquad \partial_\mu (U^\dag)^\mu=0 \label{zeroT} \end{equation}
should be satisfied. These conditions exclude a spin-0 particle.

Equation (\ref{ProcaII}) is the first Proca equation \cite{Pr}. The second Proca equation can be derived from the Lagrangian (\ref{LagrangianBc}) and takes the form \cite{Pr} \begin{equation} \partial^\nu U_{\mu\nu}-m^2 U_\mu=0.
\label{eqCorShvc}
\end{equation} The Proca functions, $U_{\mu\nu}$ and $U_\mu$, have ten independent components.

Equations (\ref{LagrangianBc}) -- (\ref{eqCorShvc}) describe a vector particle in vacuum.

In the case of the free Maxwell field $A_\mu$, both the Lagrangian density (\ref{Lagremf}) and the Maxwell equations are invariant under the local gauge transformation (\ref{gaugetf}). This is not the case for the free Proca field, where both the Lagrangian density and the Proca equations are not invariant under the local gauge transformation \begin{equation} \Psi\rightarrow\Psi'=\exp{(ie\Lambda)}\Psi,\qquad U_\mu\rightarrow U'_\mu
= U_\mu-\partial _\mu\Lambda. \label{gtrn} \end{equation}
The gauge invariance is broken by the mass term. This result is quite natural. It would be more natural to assume that the Proca Lagrangian can be gauge invariant under the local gauge transformation (\ref{gaugetf}). However, it is not gauge invariant due to the mass term. This circumstance results from the fact that the Abelian Proca model has second-class constraints. The first-class constraints are defined as the constraints which commute (i.e., have vanishing Poisson brackets) with all other constraints. This situation brings to light the presence of some
gauge degrees of freedom in the Dirac formalism. On the other hand, the second-class
constraints have at least one nonvanishing bracket with some other constraints. Models with the second-class
constraints can be converted into gauge theories with first-class constraints. The quantization of a second-class constrained system can be achieved by the reformulation of the original theory as a first-class one and then quantizing the resulting first-class
theory. Such a procedure has been performed for the Proca field \cite{Dayi,Bizdadea,Vytheeswaran,VytheeswaranBook,VytheeswaranHidden,Sararu,Abreu}. In Refs. \cite{Dayi,Bizdadea}, the BRST
quantization method has been used. It is equivalent to the BFV method \cite{BFV} for the considered problem (see Refs. \cite{Dayi,Bizdadea}). Another possibility of gauge transformations for the Proca field is an application of the method of gauge unfixing \cite{Vytheeswaran,VytheeswaranBook,VytheeswaranHidden,Sararu,Abreu}. These two methods advocate both the gauge transformation (\ref{gtrn}) and lengthening the derivatives (\ref{Dmu}) for the Proca particle in electromagnetic fields. The same conclusion also follows from the results obtained in Refs. \cite{Aldaya,Anastasovski,DarabiNaderi}. A unified quantization of both the electromagnetic and Proca fields has been performed in Ref. \cite{Aldaya}. In Ref. \cite{Anastasovski}, an elimination of the Lorenz condition has been applied. In Ref. \cite{DarabiNaderi}, a noncommutative spacetime has been used.

Besides lengthening the derivatives, the Proca Lagrangian in electromagnetic fields is usually supplemented by the Corben-Schwinger term
\cite{CS},
\begin{equation} {\cal L}_{AMM}=\frac{ie\kappa}{2}\left(U_{\mu}^\dag U_\nu-U_{\nu}^\dag U_\mu\right)F^{\mu\nu}, \label{LagranganAMM} \end{equation}
where $F^{\mu\nu}$ is the electromagnetic field tensor. The Corben-Schwinger term is proportional to $\kappa=g-1$, where $g=2m\mu/(e\hbar s)$. For
spin-1 particles, $g=2m\mu/(e\hbar)$. Since the initial Proca equations correspond to $g=1$,
this term describes not only the AMM but also a part of the normal
($g=2$) magnetic moment $\mu_0=e\hbar/m$.

The corresponding initial Proca-Corben-Schwinger equations (generalized Proca equations) describing the electromagnetic interactions of the spin-1 particle in the Minkowski spacetime have the form (see Refs. \cite{CS,YB,Spinunit})
\begin{equation} U_{\mu\nu}=D_\mu U_\nu-D_\nu U_\mu,
\label{eqProca}
\end{equation}
\begin{equation} D^\nu U_{\mu\nu}-m^2 U_\mu-ie\kappa U^\nu F_{\mu\nu}=0,~~~ F_{\mu\nu}=\partial_\mu A_\nu-\partial_\nu A_\mu,
\label{eqCorSh}
\end{equation}
where the covariant derivative $D_\mu$ has the form (\ref{Dmu}). We should mention that properties of the four-potential $A_\mu=(\Phi,-\bm A)$ in special relativity and general relativity substantially differ. In particular, covariant and contravariant components of the four-potential have even different dimensions when the metric tensor is not dimensionless.

Since the Proca Lagrangian is not gauge invariant, its extension on curved spacetimes cannot be as straightforward as for a Dirac particle. Therefore, the Proca Lagrangian in curved spacetimes has been constructed in Refs. \cite{Seitz,Spinosa,Bagrov} by a replacement of the partial derivative by the standard covariant derivative of general relativity. The same replacement is fulfilled for a scalar boson \cite{GribPob,Faraoni}. The standard covariant derivatives of the scalar $\phi$ and the covariant vector $J_\nu$ are given by
$$\begin{array}{c}
\mathfrak{D}_\mu\phi\equiv\phi_{;\mu}=\partial_\mu\phi,\qquad
\mathfrak{D}_\mu J_\nu\equiv J_{\nu\,;\mu}=\partial_\mu J_{\nu}-\left\{^{\,\rho}_{\nu\mu}\right\}J_{\rho},\end{array}$$
where
\begin{equation}
\begin{array}{c}
\left\{^{\,\rho}_{\nu\mu}\right\}=\frac12g^{\rho\lambda}\left(g_{\lambda\nu,\mu}+g_{\lambda\mu,\nu}-g_{\nu\mu,\lambda}\right)
\end{array} \label{Chrcgen} \end{equation} are the Christoffel symbols. The covariant derivative which includes the electromagnetic interactions has the form
\begin{equation} D_\mu=\mathfrak{D}_\mu+ieA_\mu. \label{covderivgr} \end{equation} We underline a substantial difference between the definitions of covariant derivatives for the Dirac and Proca particles.

In Ref. \cite{Bagrov}, the presented Proca Lagrangian includes the Corben-Schwinger term. The corresponding Proca-Corben-Schwinger (PCS) equations are given by Eqs. (\ref{eqProca}) and (\ref{eqCorSh}) where the covariant derivatives are defined by Eq. (\ref{covderivgr}).

\section{Generalized Proca equations in the Minkowski spacetime} \label{GenrProcaeqn}

Previous developments of the Proca QM including the FW transformation have been performed in Refs. \cite{YB,SpinunitEDM,PRDexact,Spinunit}. A possibility of dual transformations,  $\bm B\rightarrow \bm E,~ \bm E\rightarrow-\bm B,~\mu'\rightarrow d$, allows one to supply the Proca Lagrangian by the term characterizing the EDM \cite{SpinunitEDM},
\begin{equation} {\cal L}_{EDM}=-\frac{ie\eta}{2}\left(U_{\mu}^\dag U_\nu-U_{\nu}^\dag U_\mu\right)G^{\mu\nu}, \label{LagranganEDM} \end{equation}
where the tensor $G_{\mu\nu}=(\bm B,-\bm E)$ is dual to the electromagnetic field one, $F_{\mu\nu}=(\bm E,\bm B)$. Here $\eta=2mcd/(e\hbar s)=2mcd/(e\hbar)$, and $d$ is the EDM.
The second PCS equation takes the form \cite{SpinunitEDM}
\begin{equation} D^\nu U_{\mu\nu}-m^2 U_\mu-ie\kappa U^\nu F_{\mu\nu}+ie\eta U^\nu
G_{\mu\nu}=0.
\label{eqCorShEDM}
\end{equation}

The PCS equations can be presented in a Hamiltonian form. Since the spin of a Proca particle has three components, six
components of the wave function are independent. Spatial
components of Eq. (\ref{eqProca}) and a time component of Eqs.
(\ref{eqCorSh}) and (\ref{eqCorShEDM}) can be expressed in terms of the others. As a
result, the equations for the ten-component wave function can be
reduced to the equation for the six-component one (Sakata-Taketani
transformation \cite{SaTa}). The distinctive feature of this
transformation is that one obtains expressions for $U_0$ and
$U_{ij}~(i,j=1,2,3)$, which do not contain the time derivative, and
then substitutes them into equations for the remaining
components. From Eq. (\ref{eqCorSh}), we have
$$ U_0=\frac{1}{m^2}\left( D^i U_{0i}-ie\kappa U^i F_{0i}\right).$$
Next we introduce two vector
functions, $\bm\phi$ and $\bm U$,
the components of which are given by $iU_{i0}/m$ and $U^i$: $$ \bm\phi\equiv\frac{i}{m}\left(U_{i0}\right),
\qquad \bm U\equiv\left(U^{i}\right)=-\left(U_{i}\right).$$
We assume that the components of the vector $\bm D$ are equal to $D_i$. With these denotations,
$$ U_0=-\frac{i}{m}\bm D\cdot\bm\phi-\frac{ie\kappa}{m^2}\bm E\cdot\bm U,
\qquad \bm D\times(\bm D\times\bm U)=\left(D^jU_{ij}\right).$$ It should be underlined that there is not any difference between upper and lower components of vectors.

To perform the general Sakata-Taketani (ST) transformation of Eqs. (\ref{eqProca}) and (\ref{eqCorSh}), it is convenient to define the spin-1 matrices as follows \cite{YB}:
\begin{equation}
S^{(1)}=\left(\begin{array}{ccc} 0& 0& 0 \\ 0& 0 & -i \\ 0 & i & 0 \end{array}\right), ~~~
S^{(2)}=\left(\begin{array}{ccc} 0& 0& i \\ 0& 0 & 0 \\ -i & 0 & 0 \end{array}\right), ~~~
S^{(3)}=\left(\begin{array}{ccc} 0& -i & 0 \\ i & 0 & 0 \\ 0 & 0 & 0 \end{array}\right).
\label{spinunitmatr}
\end{equation}
This definition is not unique. One can use any other spin matrices satisfying the properties
\begin{equation}
[S^{(i)},S^{(j)}]=ie_{ijk}S^{(k)}, ~~~
S^{(i)}S^{(j)}S^{(k)}+S^{(k)}S^{(j)}S^{(i)}=\delta_{ij}S^{(k)}+\delta_{jk}S^{(i)},~~~\bm S^2=2{\cal
I},
\label{spinmatrprop}
\end{equation} where ${\cal I}$ is the unit $3\times3$ matrix.

An exclusion of the components $U_0$ and
$U_{ij}$ results in
\begin{equation}\begin{array}{c}  iD_0\bm\phi=m\bm U
+\frac{1}{m}\bm D\times(\bm D\times\bm
U)-\frac{ie\kappa}{m}\bm B\times\bm
U+\frac{e\kappa}{m^2}\bm E(\bm D\cdot\bm\phi)+\frac{e^2\kappa^2}{m^3}\bm E(\bm E\cdot\bm U),\\
iD_0\bm U=m\bm\phi-\frac{1}{m}\bm D(\bm D\cdot\bm\phi)-\frac{e\kappa}{m^2}\bm D(\bm E\cdot\bm U). \end{array}\label{eqUii}\end{equation}

The following properties are valid for any operators $\bm V$ and $\bm W$ proportional to the unit matrix ${\cal I}$:
\begin{equation}\begin{array}{c}  \bm V\times\bm\phi=-i(\bm S\cdot\bm V)\bm\phi, ~~~
\bm V(\bm W\cdot\bm\phi)=\left[\bm V\cdot\bm W-S^{(i)}S^{(j)}V^{(j)}W^{(i)}\right]\bm\phi, \\
V^{(j)}W^{(i)}\phi^{(j)}=\left[\bm V\cdot\bm W-(\bm S\cdot\bm V)(\bm S\cdot\bm W)\right]\phi^{(i)}. \end{array}\label{AandB}\end{equation}
In particular, \begin{equation}\bm D\times(\bm D\times\bm U)=-(\bm S\cdot\bm D)^2\bm U.
\label{additionl}\end{equation} Since $[D_i,D_j]=-iee_{ijk}B^{(k)}$, Eq. (\ref{eqUii}) takes the form
\begin{equation}\begin{array}{c}  iD_0\bm\phi=m\bm U    
-\frac{1}{m}(\bm S\cdot\bm D)^2\bm
U-\frac{e\kappa}{m}(\bm S\cdot\bm B)\bm
U-\frac{e\kappa}{m^2}\left[S^{(i)}S^{(j)}E^{(j)}D^{(i)}-\bm E\cdot\bm D\right]\bm\phi\\-\frac{e^2\kappa^2}{m^3}\left[(\bm S\cdot\bm E)^2-\bm E^2\right],\\
iD_0\bm U=m\bm\phi   
+\frac{1}{m}\left[(\bm S\cdot\bm D)^2-\bm D^2\right]\bm\phi
-\frac{e}{m}(\bm S\cdot\bm B)\bm\phi+\frac{e\kappa}{m^2}\left[S^{(i)}S^{(j)}D^{(j)}E^{(i)}-\bm D\cdot\bm E\right]\bm U. \end{array}\label{eqUff}\end{equation}

The wave functions $\bm\phi$ and $\bm\chi$ form the
six-component ST wave function
$$\Psi =\frac{1}{\sqrt2}\left(\begin{array}{c} \bm\phi+ \bm U \\
\bm\phi-\bm U \end{array}\right).$$
The final equation in the ST representation has the Hamiltonian form:
\begin{equation} i\frac{\partial\Psi}{\partial t}={\cal H}\Psi. \label{HamSTeq}\end{equation}
The general ST Hamiltonian obtained by Young and Bludman
\cite{YB} is given by
\begin{equation}\begin{array}{c}
{\cal H}=e\Phi+\rho_3 m+i\rho_2\frac{1}{m}(\bm S\cdot\bm
D)^2\\-(\rho_3+i\rho_2) \frac{1}{2m}(\bm D^2+e\bm S\cdot\bm B)-
(\rho_3-i\rho_2) \frac{e\kappa}{2m}(\bm S\cdot\bm B)\\-
\frac{e\kappa}{2m^2}(1+\rho_1)\biggl[(\bm S\cdot\bm E)(\bm
S\cdot\bm D)-i \bm S\cdot[\bm E\times\bm D]-\bm E\!\cdot\!\bm
D\biggr]\\ +\frac{e\kappa}{2m^2}(1-\rho_1)\biggl[(\bm S\cdot\bm
D)(\bm S\cdot\bm E)-i \bm S\cdot[\bm D\times\bm E]-\bm
D\!\cdot\!\bm
E\biggr]\\
-\frac{e^2\kappa^2}{2m^3}(\rho_3-i\rho_2)\biggl[(\bm S\cdot\bm
E)^2- \bm E^2\biggr],
\end{array} \label{eq15spinunit} \end{equation}
where $\rho_i~(i=1,2,3)$
are the $2\times2$ Pauli matrices. We do not consider a nonintrinsic quadrupole moment
included in Ref. \cite{YB}.

Equations (\ref{HamSTeq}) and (\ref{eq15spinunit}) define the general Hamiltonian form of the initial PCS equations (\ref{eqProca}) and
(\ref{eqCorSh}). For spin-1 particles, the
polarization operator is equal to $\bm\Pi=\rho_3\bm S$. It is
analogous to the corresponding Dirac operator, which can be written
in a similar form: $\bm\Pi=\rho_3\bm\sigma$.

This approach has been applied for a description of electromagnetic interactions of a Proca particle
with the EDM \cite{SpinunitEDM}. In particular, the equation of spin motion containing terms with electric and magnetic dipole moments perfectly agrees with the corresponding equations in QM of a spin-1/2 particle \cite{RPJ,RPJSTAB} and in classical physics (see Ref. \cite{classEDM} and references therein). The general description of spin motion of a Proca particle includes spin-tensor interactions proportional to terms bilinear in spin \cite{Spinunit,Bar3,BaryshevskySilenko,PRC}.

\section{Covariant Proca equations in Riemannian spacetimes} \label{CovariantProcaeqn}

The general covariant Proca equations in Riemannian spacetimes are also given by the formulas (\ref{eqProca}) and (\ref{eqCorShEDM}). However, the covariant derivatives are defined by Eq. (\ref{covderivgr}).
Next derivations can be fulfilled similarly to the case considered in the previous section. We should remind the reader of the substantial difference between the covariant derivatives for the Dirac and Proca particles. The covariant derivative contains the spin-dependent term for the Dirac particle [see Eqs. (\ref{Lorcovdv}) and (\ref{Lorcopem})] but cannot contain such a term for the Proca particle. The spin matrices for Proca particles naturally appear as a result of a transition from the ten-component wave function to the
six-component one \cite{YB,Spinunit}. Therefore, they cannot act on the Proca fields $U_{\mu\nu}$ and $U_\mu$.

It is necessary to specify that the spin is defined in the particle rest frame which belongs to local Lorentz (tetrad) frames. Therefore, the spin matrices for the Dirac and Proca particles have the standard form only in local Lorentz frames (LLFs). As a result, the spin matrix $\bm S$ should be coupled with vectors defined in such frames. It is convenient to pass to vector denotations and to introduce the two vector
functions, $\bm\phi$ and $\bm U$, and the vector operator $\bm D$ \cite{YB,Spinunit,SpinunitEDM}.
In Riemannian spacetimes, they should be defined by $$\bm\phi\equiv\frac{i}{m}\left(U_{\widehat{i}\widehat{0}}\right),
\qquad \bm U\equiv\left(U^{\widehat{i}}\right),\qquad\widehat{\bm D}\equiv(D_{\widehat{i}}),$$ where $D_a=e_a^\mu D_\mu$. It is convenient to introduce the $3\times 3$ spin matrices \cite{YB,Spinunit,SpinunitEDM}. For example,
\begin{equation}\widehat{\bm D}\times(\widehat{\bm D}\times\bm U)=-(\bm S\cdot\widehat{\bm D})^2\bm U.
\label{additiont}\end{equation}

It is also convenient to pass to tetrad components in Eqs. (\ref{eqProca}) and (\ref{eqCorShEDM}). Equation (\ref{eqProca}) takes the form
\begin{equation}\begin{array}{c} U_{ab}=e_a^\mu e_b^\nu (D_\mu e_\nu^dU_d-D_\nu e_\mu^cU_c)= e_a^\mu D_\mu U_b-e_b^\nu D_\nu U_a\\+(\Gamma_{acb}-\Gamma_{bca})U^b= e_a^\mu D_\mu U_b-e_b^\nu D_\nu U_a+C_{abc}U^c,
\end{array}\label{Procagrav}
\end{equation} where the coefficients $\Gamma_{abc}$ and $C_{abc}$ are defined by Eq. (\ref{Lorcovdv}). A transformation of Eq. (\ref{eqCorShEDM}) is similar and results in
\begin{equation} e^{b\mu} D_\mu U_{ab}-\Gamma_{bca}U^{bc}+\eta^{bc}\eta^{df}C_{cdf}U_{ab}-
m^2 U_a-ie\kappa U^b F_{ab}+ie\eta U^b G_{ab}=0,
\label{PrCorSh} \end{equation}
where $F_{ab}=e^{a}_{\mu}e^{b}_{\nu}F_{\mu\nu},~G_{ab}=e^{a}_{\mu}e^{b}_{\nu}G_{\mu\nu}$.

The general equations (\ref{Procagrav}) and (\ref{PrCorSh}) are the
covariant equations describing electromagnetic interactions of a Proca particle with the AMM and EDM in Riemannian spacetimes.
The presence of the Lorentz connection coefficients in Eqs. (\ref{Procagrav}) and (\ref{PrCorSh}) leaves room for effects caused by the Cartan torsion.

In Refs. \cite{gravity,PRDThomas,PK}, a similarity between $(e/m)F_{\mu\nu}$ and $\Gamma_{ceb}u^c$ has been stated. In this connection, it apparently seems that the definition of the Proca fields $U_{ab},~U_{a}$ in LLFs allows one to include the additional term $-im\upsilon U^b \Gamma_{ceb}u^c~(\upsilon=const)$ into
Eq. (\ref{PrCorSh}). However, the tensorlike quantity $e_\mu^e e_\nu^b\Gamma_{ceb}u^c =e_\mu^ee_{e\nu;\rho}u^\rho$ (unlike $F_{\mu\nu}$) is anholonomic. Therefore, it is not a true tensor, and the additional term $-im\upsilon U^\nu e_\mu^e e_\nu^b\Gamma_{ceb}u^c$ cannot enter Eq.
(\ref{eqCorShEDM}), which is fully covariant.

For an example, we derive the FW Hamiltonian, taking into account only terms proportional to the zero power of $\hbar$. These terms define the particle motion. A derivation of smaller terms describing the spin motion needs rather cumbersome calculations.

\section{Foldy-Wouthuysen transformation for a Proca particle in Riemannian spacetimes} \label{FWtProca}

Since we disregard terms of the first and higher orders in $\hbar$, we can neglect the terms describing the AMM and EDM and all terms originating from commutators with the operator $D_\mu$. In this approximation, Eqs. (\ref{Procagrav}) and (\ref{PrCorSh}) reduce to
\begin{equation}\begin{array}{c} U_{ab}= e_a^\mu D_\mu U_b-e_b^\nu D_\nu U_a,\qquad e^{b\mu} D_\mu U_{ab}-
m^2 U_a=0.
\end{array} \label{PrCorShlittl} \end{equation}
In further derivations, commutators of the operator $D_\mu$ with other operators entering Eq. (\ref{PrCorShlittl}) can also be neglected.

The previous analysis shows 
that the best choice is the Schwinger gauge (see Refs. \cite{gravity,PRDThomas})
satisfying the relations $e_i^{\,\widehat{0}} =0,~e^0_{\,\widehat{i}} = 0$. In this case, the operator $D_{\widehat{i}}=e_{\widehat{i}}^jD_{j}$ does not contain $D_{0}$. Of course, other gauges can also be used, 
but they are much less convenient. For the Schwinger gauge, \begin{equation}g^{00}=e_{\widehat{0}}^0e^{\widehat{0}0}=\left(e_{\widehat{0}}^0\right)^2,\qquad
g^{0i}=e^{\widehat{0}0}e_{\widehat{0}}^{i}=e_{\widehat{0}}^{0}e_{\widehat{0}}^{i}.\label{gnulnul}\end{equation}
Another important property of this gauge is valid for any covariant operator:
\begin{equation}\mathcal{B}_i=e^{\widehat{j}}_i\mathcal{B}_{\widehat{j}},\qquad \mathcal{B}_{\widehat{i}}=e_{\widehat{i}}^j\mathcal{B}_{j}.\label{acovope}\end{equation}
Equations (\ref{gnulnul}) and (\ref{acovope}) are valid for any Schwinger gauge.

We follow the same approach as that applied in Refs. \cite{Spinunit,SpinunitEDM}. To eliminate the components $U_{\widehat{0}}$ and
$U_{\widehat{i}\widehat{j}}$, we obtain expressions for them and substitute these expressions into equations for the remaining components. For the introduced vectors,
\begin{equation}\widehat{\bm D}\cdot\bm\phi\equiv -\frac{i}{m}\eta^{\widehat{i}\widehat{j}}D_{\widehat{i}}U_{\widehat{i}\widehat{0}},\qquad\widehat{\bm D}\cdot\bm U\equiv D_{\widehat{i}}U^{\widehat{i}},\qquad
\bm {\widehat{D}}^2\equiv-\eta^{\widehat{i}\widehat{j}}D_{\widehat{i}}D_{\widehat{j}}. \label{Dquae} \end{equation}
The eliminated components are given by 
\begin{equation}\begin{array}{c} U_{\widehat{i}\widehat{j}}= D_{\widehat{i}} U_{\widehat{j}}-D_{\widehat{j}} U_{\widehat{i}},\qquad U_{\widehat{0}}=-\frac{i}{m}\widehat{\bm D}\cdot\bm\phi.
\end{array} \label{PrCorSchw} \end{equation}

Next derivations are similar to those fulfilled in Refs. \cite{Spinunit,SpinunitEDM}. In the approximation used, we obtain the following equations for the wave functions $\bm\phi$ and $\bm U$:
\begin{equation}\begin{array}{c}  iD_{\widehat{0}}\bm\phi=m\bm U
+\frac{1}{m}\widehat{\bm D}\times(\widehat{\bm D}\times\bm
U),\\
iD_{\widehat{0}}\bm U=m\bm\phi-\frac{1}{m}\widehat{\bm D}(\widehat{\bm D}\cdot\bm\phi). \end{array}\label{eqUiigr}\end{equation}

An introduction of the spin matrices brings Eq. (\ref{eqUiigr}) to the form
\begin{equation}\begin{array}{c}  iD_{\widehat{0}}\bm\phi=m\bm U    
-\frac{1}{m}(\bm S\cdot\widehat{\bm D})^2\bm
U,\\
iD_{\widehat{0}}\bm U=m\bm\phi   
+\frac{1}{m}\left[(\bm S\cdot\widehat{\bm D})^2-{\widehat{\bm D}}^2\right]\bm\phi. \end{array}\label{eqUffgr}\end{equation}

Similarly to Refs. \cite{Spinunit,SpinunitEDM}, the wave functions $\bm\phi$ and $\bm U$ form the
six-component Sakata-Taketani wave function
$$\Psi =\frac{1}{\sqrt2}\left(\begin{array}{c} \bm\phi+ \bm U \\
\bm\phi-\bm U \end{array}\right).$$
The resulting equation in the Sakata-Taketani representation has the form
\begin{equation} iD_{\widehat{0}}\Psi=\left[\rho_3 m+i\rho_2\frac{1}{m}\left(\bm S\cdot\widehat{\bm D}\right)^2-(\rho_3+i\rho_2) \frac{1}{2m}{\widehat{\bm D}}^2\right]\Psi, \label{HamSTgr}\end{equation}
where $\rho_i$ are the $2\times2$ Pauli matrices.

We need to bring this equation to the Hamiltonian form and then to perform the FW transformation. Since
$$D_{\widehat{0}}=e_{\widehat{0}}^0D_0+e_{\widehat{0}}^iD_i,$$ the use of Eq. (\ref{gnulnul}) allows us to determine the operator $D_0$:
\begin{equation}
D_0=\frac{1}{\sqrt{g^{00}}}D_{\widehat{0}}-\frac{g^{0i}}{g^{00}}D_i.
\label{Dworl}\end{equation}

The wave functions $\bm\phi,\,\bm U$, and $\Psi$ have tetrad components but not world ones. Therefore, their first covariant derivatives are similar to covariant derivatives of a scalar wave function and $D_\mu\Psi=(\partial_\mu+ieA_\mu)\Psi,~D_a\Psi=e_a^\mu D_\mu\Psi$.

As a result, the Hamiltonian form of Eq. (\ref{HamSTgr}) is given by
\begin{equation}\begin{array}{c} i\frac{\partial\Psi}{\partial t}= iD_{0}\Psi={\cal H}\Psi,\\
{\cal H}=\frac{1}{\sqrt{g^{00}}}\left[\rho_3 m+i\rho_2\frac{1}{m}\left(\bm S\cdot\widehat{\bm D}\right)^2-(\rho_3+i\rho_2) \frac{1}{2m}{\widehat{\bm D}}^2\right]-\frac{g^{0i}}{g^{00}}iD_i+eA_0. \end{array}\label{HamSTtr}\end{equation}

We can now perform the FW transformation by the method described in Ref. \cite{relativisticFW}. Any initial Hamiltonan can be presented in the form
\begin{equation} {\cal H}=\beta{\cal M}+{\cal E}+{\cal
O},~~~\beta{\cal M}={\cal M}\beta, ~~~\beta{\cal E}={\cal E}\beta, ~~~\beta{\cal O}=-{\cal
O}\beta, \label{eq3} \end{equation} where the operators ${\cal M}$ and ${\cal E}$ are even and the operator ${\cal
O}$ is odd. The matrix $\beta$ is the direct product of the Pauli matrix $\rho_3$ and the $3\times 3$ unit matrix ${\cal I}$. Even and odd operators are diagonal and off-diagonal in two spinorlike parts of the bispinorlike wave function $\Psi$ and commute and anticommute with the operator $\beta$, respectively. Explicitly,
\begin{equation}\begin{array}{c} {\cal M}=\frac{1}{\sqrt{g^{00}}}\left(m-\frac{1}{2m}{\widehat{\bm D}}^2\right), ~~~ {\cal E}=-\frac{g^{0i}}{g^{00}}iD_i+eA_0,\\ {\cal
O}=\frac{i\rho_2}{\sqrt{g^{00}}m}\left[\left(\bm S\cdot\widehat{\bm D}\right)^2-\frac{1}{2}{\widehat{\bm D}}^2\right]. \end{array}\label{MEO}\end{equation}

The noncommutativity of the operator $D_\mu$ with the metric tensor and with the tetrad leads to the appearance of terms proportional to $\hbar$. Since we neglect such terms, we can ignore the above-mentioned noncommutativity, and the transformed Hamiltonian takes the form
\begin{equation}\begin{array}{c}
{\cal H}_{FW}=\frac{\rho_3}{\sqrt{g^{00}}}\sqrt{m^2-{\widehat{\bm D}}^2}-\frac{g^{0i}}{g^{00}}iD_i+eA_0. \end{array}\label{HamFW}\end{equation}

The use of Eqs. (\ref{gnulnul}) and (\ref{acovope}) allows one to express the operator ${\widehat{\bm D}}^2$ in terms of covariant derivatives. For any tetrad belonging to the Schwinger gauge, the following relation is valid:
\begin{equation}g^{ij}=e^i_{\widehat{0}}e^{\widehat{0}j}+e^i_{\widehat{k}}e^{\widehat{k}j}=\frac{g^{0i}g^{0j}}{g^{00}}+e^i_{\widehat{k}}e^{\widehat{k}j}.\label{gtetr}\end{equation}
Equations (\ref{acovope}) and (\ref{gtetr}) show that in the approximation used
\begin{equation}G^{ij}D_{i}D_{j}\equiv\left(g^{ij}-\frac{g^{0i}g^{0j}}{g^{00}}\right)D_{i}D_{j}=e^i_{\widehat{k}}e^{\widehat{k}j}D_{i}D_{j}=
D_{\widehat{k}}D^{\widehat{k}}\equiv{\widehat{\bm D}}^2 \label{gtetn}\end{equation}
and
\begin{equation}\begin{array}{c}
{\cal H}_{FW}=\frac{\rho_3}{\sqrt{g^{00}}}\sqrt{m^2-G^{ij}D_{i}D_{j}}-\frac{g^{0i}}{g^{00}}iD_i+eA_0. \end{array}\label{HamFWfn}\end{equation}

Since covariant derivatives of the metric tensor are equal to zero, $G^{ij}$ commutes with $D_\mu$.

Evidently, the Hamiltonian (\ref{HamFWfn}) agrees with the corresponding classical Hamiltonian which has the form (Eq. (2.5) in Ref. \cite{Cogn})
\begin{equation}
H = \left(\frac{m^2 - G^{ij}p_ip_j} {g^{00}}\right)^{1/2} -
\frac{g^{0i}p_i}{{g}^{00}}, \qquad G^{ij}=g^{ij}-\frac{g^{0i}g^{0j}}{g^{00}}. \label{clCog}
\end{equation}

The Hamiltonian (\ref{HamFWfn}) covers the electromagnetic and gravitational
interactions. It also describes the inertial interactions taking place in flat noninertial frames. We should remind the reader that the Hamiltonian  (\ref{HamFWfn}) acts on the six-component wave function and the unit matrices are omitted. The agreement of this Hamiltonian with the contemporary QM is shown in the next section.

\section{Comparison of Foldy-Wouthuysen Hamiltonians in Minkowski and Riemann spacetimes} \label{Comparison}

It is instructive to compare the result obtained in the present work with the contemporary Proca QM in the Minkowski spacetime. The comparison is nontrivial when curvilinear coordinates are used. The momentum operator is proportional to the nabla one, $\bm p\equiv-(p_i)=-i\hbar(\nabla_i)\equiv-i\hbar\nabla$. For particles with any spin, the terms in the FW Hamiltonians proportional to the zero power of the Planck constant are given by \cite{RPJ,TMP2008,SpinunitEDM}
\begin{equation}\begin{array}{c}
{\cal H}_{FW}=\rho_3\sqrt{m^2+\bm\pi^2}+e\mathcal{A}_0,\qquad \bm\pi=\bm p-e\bm{\mathcal{A}}. \end{array}\label{HamFWMs}\end{equation}
These terms are independent of the spin.

The four-potentials $A_\mu$ and $\mathcal{A}_\mu$ defining the same electromagnetic field in the flat Riemannian spacetimes (when curvilinear coordinates are used) and in the Minkowski spacetime, respectively, substantially differ. For any diagonal metric tensor, the appropriate choice of the Schwinger tetrad is $e_a^\mu=\delta_{a\nu}\sqrt{|g^{\mu\nu}|}$.
In this case, $\widehat{\bm A}=\bm{\mathcal{A}}$ but $\bm A\neq\bm{\mathcal{A}}$.

It is well known that the nabla operator possesses nontrivial properties even in the Minkowski spacetime. It is defined by
\begin{equation}
\nabla=\frac{\partial}{\partial\rho}\bm e_\rho+\frac{1}{\rho}\frac{\partial}{\partial\phi}\bm e_\phi+\frac{\partial}{\partial z}\bm e_z
\label{nablacyl} \end{equation} and \begin{equation}
\nabla=\frac{\partial}{\partial r}\bm e_r+\frac{1}{r}\frac{\partial}{\partial\theta}\bm e_\theta+\frac{1}{r\sin{\theta}}\frac{\partial}{\partial\phi}\bm e_\phi
\label{nablasph} \end{equation}
for the cylindrical $(\rho,\phi,z)$ and spherical $(r,\theta,\phi)$ coordinates, respectively. Its action on a scalar is trivial, but its convolution with a vector (divergence of the vector) in these coordinates is nontrivial and has the form
\begin{equation}\begin{array}{c}
\nabla\cdot{\bm{\mathfrak{T}}}=\frac{1}{\rho}\frac{\partial(\rho{\mathfrak{T}}_\rho)}{\partial\rho}+\frac{1}{\rho}\frac{\partial{\mathfrak{T}}_\phi}{\partial\phi}+\frac{\partial{\mathfrak{T}}_z}{\partial z},\\
\nabla\cdot{\bm{\mathfrak{T}}}=\frac{1}{r^2}\frac{\partial(r^2{\mathfrak{T}}_r)}{\partial r}+\frac{1}{r\sin{\theta}}\frac{\partial(\sin{\theta}{\mathfrak{T}}_\theta)}{\partial\theta}+\frac{1}{r\sin{\theta}}\frac{\partial{\mathfrak{T}}_\phi}{\partial\phi}.
\end{array} \label{nablavec} \end{equation} In this equation, $\bm{\mathfrak{T}}$ is a three-component vector and there is not any difference between its covariant and contravariant components.

The operator $\bm p^2=-\hbar^2\Delta$ is proportional to the Laplace operator $\Delta\equiv\nabla\cdot\nabla$. It acts on the scalar wave function. This operator is defined by
\begin{equation}\begin{array}{c}
\Delta=\frac{1}{\rho}\frac{\partial}{\partial\rho}\left(\rho\frac{\partial}{\partial\rho}\right)+\frac{1}{\rho^2}\frac{\partial^2}{\partial\phi^2}+\frac{\partial^2}{\partial z^2} \end{array} \label{Deltaspher} \end{equation} and \begin{equation}\begin{array}{c}
\Delta=\frac{1}{r^2}\frac{\partial}{\partial r}\left(r^2\frac{\partial}{\partial r}\right)+\frac{1}{r^2\sin{\theta}}\frac{\partial}{\partial\theta}\left(\sin{\theta}\frac{\partial}{\partial\theta}\right)+\frac{1}{r^2\sin^2{\theta}}\frac{\partial^2}{\partial\phi^2}
\end{array} \label{Deltacyl} \end{equation}
in the cylindrical and spherical coordinates, respectively.

We can mention that
$$\mathfrak{T}^{;\mu}_\mu=\frac{1}{\sqrt{-g}}\partial_\mu\sqrt{-g}g^{\mu\nu}\mathfrak{T}_\nu.$$
For any curvilinear coordinates, $g_{00}=1$ and the metric is static ($g_{0i}=0$). In this case,
$$\mathfrak{T}^{;i}_i=\frac{1}{\sqrt{-g}}\partial_i(\sqrt{-g}g^{ij}\mathfrak{T}_j).$$

For the cylindrical and spherical coordinate systems, the metric tensor is defined by $g_{\mu\nu}={\rm diag}(1, -1, -\rho^2, -1)$ and $g_{\mu\nu}={\rm diag}(1, -1,-r^2, -r^2\sin^2{\theta})$, respectively. It can be shown that the operators
$G^{ij}D_iD_j$ and $-(\nabla-ie\bm A)\cdot(\nabla-ie\bm A)$ are equivalent
in the both cases. In these cases, $G^{ij}=g^{ij}$. First, it can be easily checked that the operators $G^{ij}D_iD_j$ and $-\Delta\equiv-\nabla\cdot\nabla$ are equivalent. Second, it is necessary to consider the divergence $G^{ij}A_{j\,;i}=A^{;i}_i$. For the cylindrical coordinates [$\bm{\mathcal{A}}=-({\mathcal{A}}_i)$], $$-A^{;i}_i=\frac1\rho\partial_\rho(\rho A_\rho)+\frac{1}{\rho^2}\partial_\phi A_\phi+\partial_z A_z=\frac1\rho\partial_\rho(\rho A_{\hat{\rho}})+\frac{1}{\rho}\partial_\phi A_{\hat{\phi}}+\partial_z A_{\hat{z}}=-\nabla\cdot\widehat{\bm A}=-\nabla\cdot\bm{\mathcal{A}}.$$
The same result can be obtained for the spherical coordinates.

Since $G^{ij}A_iA_j=A^aA_a=-\widehat{\bm A}^2=-\bm{\mathcal{A}}^2$, the equivalence of the operators $G^{ij}D_iD_j$ and $-(\nabla-ie\bm A)\cdot(\nabla-ie\bm A)$ is proven.

Thus, the result obtained in the present study fully agrees with the contemporary QM in the Minkowski space.

\section{Unification and classical limit of relativistic quantum mechanics in the Foldy-Wouthuysen representation} \label{Unification}

The Hamiltonian (\ref{HamFWfn}) also fully agrees with the corresponding FW Hamiltonians for a scalar particle \cite{Honnefscalar} and for a Dirac one \cite{gravity}. In the latter Hamiltonians, we can disregard terms of the first and higher orders in the Planck constant.
The Hamiltonians differ only in the dimensions of
contained matrices defined by the dimensions of the corresponding wave functions. For states with a positive total energy, lower spinors (or lower parts of spinorlike wave functions) are equal to zero for any particles. Their nullification unifies a normalization of the wave functions. For any spin, the FW wave functions are normalized to unit, and their probabilistic interpretation is restored. It should be underlined that the quantum-mechanical Hamiltonians become rather similar for bosons and fermions. A difference between the Hermiticity of the initial Hamiltonians for fermions \cite{gravity} and the $\beta$-pseudo-Hermiticity of the corresponding Hamiltonians for bosons \cite{Honnefscalar} disappears after the FW transformation. These properties indicate the unification of relativistic QM for particles with different spins in the FW representation.

In this paper, we do not analyze terms of the first order in the Planck constant. Such terms define spin interactions. However, it has been shown in Ref. \cite{SpinunitEDM} that the spin-dependent terms in the FW Hamiltonians for
spin-1/2 and spin-1 particles with the AMMs and EDMs interacting with arbitrary electromagnetic fields in Minkowski spacetimes perfectly agree. These terms define equations of spin motion, which coincide with each other in the
classical limit. These equations also coincide with the corresponding classical equations (see Ref. \cite{classEDM} and references therein). Of course, the FW Hamiltonian for spin-1 particles additionally contains bilinear in spin
terms which also influence spin dynamics \cite{Spinunit,Bar3,BaryshevskySilenko,PRC}. 

It can be concluded that the use of the FW representation allows one to unify the main equations of relativistic QM for particles with different spins and to demonstrate that their classical limit agrees with the corresponding classical equations. This conclusion fully agrees with the results obtained in Ref. \cite{PK} in which the specific quantum-mechanical approach has been used.

\section{Summary}

A comparison of fundamentals of Dirac and Proca QM shows that the problem of quantization with an introduction of
interactions can be solved more easy for a Dirac particle than for a Proca (spin-1) one. However, the solution of this problem for the Proca particle is possible \cite{Dayi,Bizdadea,Vytheeswaran,VytheeswaranBook,VytheeswaranHidden,Sararu,Abreu,Aldaya,Anastasovski,DarabiNaderi} while it meets some difficulties. A consideration of the results obtained for the Proca particle in electromagnetic fields \cite{YB,SpinunitEDM,Spinunit} demonstrates an importance of the ST and FW transformations which result in the Schr\"{o}dinger form of the PCS equations. After this, the classical limit of Proca QM in electromagnetic fields can be easily determined. Therefore, a development of Proca QM needs not only a formulation of general covariant Proca equations in electromagnetic and gravitational fields but also a determination of the Hamiltonian form and of the classical limit of these equations with the use of the ST and FW transformations. These results in turn allow one to establish a connection of QM of the Proca particle with QM of particles with other spins.

The present work proposes the extension of relativistic QM of a Proca particle on Riemannian spacetimes. The formulated covariant Proca equations take into account the AMM and the EDM of a spin-1 particle and are based on the PCS equations in special relativity and precedent studies of the Proca particle in curved spacetimes. It is important to mention that the covariant derivatives in the Dirac and Proca equations substantially differ.
As an example, the relativistic FW transformation with allowance for terms proportional
to the zero power of the Planck constant has been performed. The Hamiltonian obtained agrees with the corresponding Hamiltonians derived for scalar and Dirac particles and with their classical counterpart. This conclusion is in agreement with the results obtained in Ref. \cite{PK}. The consideration presented demonstrate the unification of relativistic QM in the FW representation.

\section*{Acknowledgments}

The author thanks F. W. Hehl for reading the preliminary
draft and for his valuable comments and discussions. The author is also grateful to an anonymous referee for important comments and propositions.
This work was supported in part by the Belarusian Republican Foundation for Fundamental
Research (Grant Nos. $\Phi$16D-004 and $\Phi$18D-002) and 
by the Heisenberg-Landau program of the German Ministry for Science and Technology (BMBF).


\end{document}